\documentclass[prd,preprintnumbers,floatfix,aps,notitlepage,nofootinbib,twocolumn,showpacs,amssymb]{revtex4-2}
\usepackage{amsmath,amsfonts,bm}
\usepackage{graphicx}
\usepackage{cancel}
\usepackage[colorlinks, linkcolor={red},citecolor={blue}]{hyperref}
\usepackage{xcolor}

\begin{document}
\title{Cosmology from Schwarzschild black hole revisited}
\author{Roberto Casadio}
\email{casadio@bo.infn.it}
\affiliation{Dipartimento di Fisica e Astronomia,
	Alma Mater Universit\`a di Bologna,
	40126 Bologna, Italy
	\\
	Istituto Nazionale di Fisica Nucleare, I.S.~FLaG
	Sezione di Bologna, 40127 Bologna, Italy}
\author{Alexander Kamenshchik}
\email{kamenshchik@bo.infn.it}
\affiliation{Dipartimento di Fisica e Astronomia,
	Alma Mater Universit\`a di Bologna,
	40126 Bologna, Italy
	\\
	Istituto Nazionale di Fisica Nucleare, I.S.~FLaG
	Sezione di Bologna, 40127 Bologna, Italy}
\author{Jorge Ovalle}
\email[]{jorge.ovalle@physics.slu.cz}
\affiliation{Research Centre for Theoretical Physics and Astrophysics,
Institute of Physics, Silesian University in Opava, CZ-746 01 Opava,
Czech Republic.}
\begin{abstract}
We study cosmological models based on the interior of the revisited
Schwarzschild black hole recently reported in [Phys.~Rev.~D{\bf 109} (2024) 104032].
We find that these solutions describe a non-trivial Kantowski-Sachs universe, for
which we provide an explicit analytical example with all the details and describe
some general features of the singularity.
\end{abstract} 
\maketitle
%
%
%
\section{Introduction}
\label{sec:intro}
The Schwarzschild metric is the most famous and simplest black hole (BH) solution in general relativity (GR),
containing a point-like singularity of ADM mass ${\cal M}$ hidden behind the event horizon or radius
$2\,{\cal M}$. 
Ref.~\cite{Ovalle:2024wtv} investigated alternative sources for the exterior region of the Schwarzschild
BH in GR, under the conditions that it does not contain any form of exotic matter nor does it depend
on any new parameter other than ${\cal M}$.
A large set of new solutions was then found that can describe the inner region before all matter energy
contributing to the total mass ${\cal M}$ has collapsed into the singularity in accordance with Penrose's
singularity theorem~\cite{Penrose:1964wq}.
\par
The most attractive features of these interiors are that the space-time across the horizon is continuous
(without additional structures like a thin shell) and tidal forces remain finite everywhere for
(integrable~\cite{Lukash:2013ts,Casadio:2021eio,Casadio:2023iqt}) singular solutions.
The importance of some of these solutions lies in the fact that they might be alternative to the
Schwarzschild BH as the final stage of the gravitational collapse and are therefore potentially
useful for studying the gravitational collapse of compact astrophysical objects: 
if we enforce the weak cosmic censorship conjecture~\cite{Penrose:1969pc} to avoid naked singularities,
the event horizon must form before the central singularity.
This means that (at least) part of the total mass ${\cal M}$ enclosed within the event horizon
is still on the way to the final singularity.
Such a scenario is precisely described by the (integrable) singular solutions reported in
Ref.~\cite{Ovalle:2024wtv}.
\par
Finally, we want to highlight that the existence of such an (incomplete) set of solutions shows
very explicitly the broad diversity that is possible in the interior region of the Schwarzschild
geometry.
This is very attractive if we want to study the appearance of singularities, not only associated 
with the formation of BHs, but also in the very early Universe. 
The main goal of this work is precisely to investigate cosmological consequences that
can be derived from the aforementioned solutions.
\section{Inside the black hole}
\label{sec:insideBH}
We begin by recalling that the most general static and spherically symmetric metric
can be written as~\cite{Morris:1988cz}
\begin{equation}
ds^{2}
=
-e^{\Phi(r)}
f(r)\,dt^{2}
+\frac{dr^2}{f(r)}
+r^2\,d\Omega^2
\ ,
\label{metric}
\end{equation}
where
\begin{equation}
f
=
1-\frac{2\,m(r)}{r}
\ .
\end{equation}
The Schwarzschild solution~\cite{Schwarzschild:1916uq} is obtained by setting
\begin{equation}
\label{KScond}
\Phi(r)
=
0
\end{equation}
and the Misner-Sharp mass function
\begin{equation}
m(r)
=
{\cal M}
\ ,
\quad
{\rm for}
\ r>0
\ ,
\label{mcond}
\end{equation}
where ${\cal M}$ is the ADM mass associated with a point-like singularity at the center $r=0$.
The coordinate singularity at $r=2\,{\cal M}\equiv\,h$ indicates the event horizon~\cite{Eddington:1924pmh,Lemaitre:1933gd,Finkelstein:1958zz,Kruskal:1959vx,Szekeres:1960gm}.
\par
We are interested in exploring extensions of the Schwarzschild BH which still belong to the
Kerr-Schild class~\cite{kerrchild} and will therefore keep the condition~\eqref{KScond} but
relax the condition~\eqref{mcond} to
\begin{equation}
m(r)={\cal M}
\ ,
\quad
{\rm for}\
r\,{\color{blue}\geq}\,h
\ ,
\end{equation}
where~\footnote{We shall denote $F(h)\equiv\,F(r)\big\rvert_{r=h}$ for any $F=F(r)$.
We shall also use units with $c=1$ and $\kappa=8\,\pi\,G_{\rm N}$.}
\begin{equation}
\label{cond1}
{\cal M}\equiv\,m(h)
=
{h}/{2}
\end{equation} 
stands for the total mass of the BH and $h$ is again the radius of its event horizon. 
In summary, the line element for the problem we want to solve is given by
\begin{equation}
\label{mtransform}
ds^2=\left\{
\begin{array}{l}
-\left(1-\frac{2\,m}{r}\right) dt^{2}
+\frac{dr^2}{1-\frac{2\,m}{r}}
+r^2\,d\Omega^2
\ ,
\
0< r\,{\color{blue}\leq}\,h
\\
\\
-\left(1-\frac{2\,{\cal M}}{r}\right)dt^{2}
+\frac{dr^2}{1-\frac{2\,{\cal M}}{r}}
+r^2\,d\Omega^2
\ ,
\
r>h
\ ,
\end{array}
\right.
\end{equation}
where $m$ is the mass function associated with a Lagrangian ${\cal L}_{\rm M}$
representing ordinary matter only, so that our theory is described by the 
Einstein-Hilbert action
\begin{equation}
\label{action}
S
=
\int\left(\frac{R}{2\,\kappa}+{\cal L}_{\rm M}\right)
\sqrt{-g}\,d^4x
\ ,
\end{equation}
with $R$ the scalar curvature.
Notice that Eqs.~\eqref{mtransform} and~\eqref{action} imply that ${\cal L}_{\rm M}=0$
for $r>h$ only, and the Einstein field equations in the interior $0<r<h$ yield an 
energy-momentum tensor 
\begin{eqnarray}
\label{emt}
T^\mu_{\ \nu}
=
{\rm diag}\left[p_r,-\epsilon,p_\theta,p_\theta\right]
\ ,
\end{eqnarray}
such that the energy density $\epsilon$, radial pressure $p_r$ and transverse pressure $p_\theta$
satisfy
\begin{eqnarray} 
\label{sources}
\epsilon=\frac{2\,{m}'}{\kappa\,r^2}
\ ,\quad
p_r=-\frac{2\,{m}'}{\kappa\,r^2}
\ ,\quad
p_\theta=-
\frac{{m}''}{\kappa\,r}
\ .
\end{eqnarray}
We remark that Eq.~\eqref{emt} takes into account the fact that the radial and temporal coordinates
exchange roles for $0<r<h$.
Finally, since Eqs.~\eqref{sources} are linear in the mass function ${m}$, 
any two solutions can be linearly combined to produce a new solution, 
which represents a trivial case of the so-called gravitational decoupling~\cite{Ovalle:2017fgl,Ovalle:2019qyi}. 
\par
The contracted Bianchi identities $\nabla_\mu\,G^{\mu}_{\,\,\nu}=0$ leads to $\nabla_\mu\,{T}^{\mu\nu}=0$, which yields
the continuity equation
\begin{eqnarray}
\label{con111}
\epsilon'=
-\frac{2}{r}\left(p_\theta-p_r\right)
\ .
\end{eqnarray}
Since the energy density $\epsilon$ is expected to decrease monotonically from the origin outwards,
that is $\epsilon'<0$, Eq.~\eqref{con111} implies that
\begin{equation}
\label{anis}
p_\theta>p_r
\ ,
\end{equation}
so that the fluid experiences a pull towards the center as a consequence of negative energy
gradients $\epsilon'<0$ that is canceled by a gravitational repulsion caused by the anisotropic
pressure.
\par
We next need to examine the compatibility of the Schwarzschild exterior with the above
interior, i.e.~the continuity of the metric~\eqref{mtransform} across the horizon $r=h$.
This clearly implies that the mass function must satisfy the matching conditions
\begin{equation}
\label{cond2}
m(h)={\cal M}
\ ,
\quad
m'(h)=0
\ .
\end{equation}
Finally, we see from Eqs.~\eqref{sources} and~\eqref{cond2} that the continuity of the mass
function implies the continuity of both density and radial pressure,
\begin{equation}
\label{c2a}
\epsilon(h)=p_r(h)=0
\ .
\end{equation}
However, the pressure $p_\theta$ can be in general discontinuous. 
\section{Black holes with\\integrable singularities}
\label{sec:integrable}
BH geometries are usually grouped into two types:
(i)~singular BH with a physical singularity of some kind, and
(ii)~regular BH without singularities.
The existence of regular BHs is, of course, very attractive but it is well known that they usually
display an inner (Cauchy) horizon inside the event horizon, which turns out to give rise to
problems such as mass inflation, instability, and eventual loss of
causality~\cite{Poisson:1989zz,Poisson:1990eh} (see also Refs.~\cite{Ori:1991zz,
Carballo-Rubio:2018pmi,Bonanno:2020fgp,Carballo-Rubio:2021bpr,
Carballo-Rubio:2022kad,Franzin:2022wai,Casadio:2022ndh,
Bonanno:2022jjp,Casadio:2023iqt} for recent studies).
Between the two aforementioned families we can also find integrable BHs~\cite{Lukash:2013ts},
which are characterized by a singularity in the curvature $R$ that occurs at most as
\begin{equation}
\label{inte-sing}
R
\sim
r^{-2}
\ ,
\end{equation}
so that their Einstein-Hilbert action is indeed finite.
Their main features are that tidal forces remain finite everywhere, the mass function is well-defined
and finite, and (in general) there are no Cauchy horizons.
\par
Regarding the last feature, we here review the work in Ref.~\cite{Ovalle:2024wtv} and
start with the scalar curvature for the interior metric~\eqref{mtransform}, 
which reads
\begin{equation}
\label{R}
R
=
\frac{2\,r\,m''+4\,m'}{r^2}
\ ,
\quad
{\rm for}\
0< r\,\,{\color{blue}\leq}\,\,h
\ .
\end{equation}
In order to have an integrable singularity, we demand~\cite{Ovalle:2024wtv} 
\begin{equation}
\label{R2}
R
=
\sum_{n=0}^{\infty}\,C_n\,r^{n-2}
\ ,
\quad
n\in\mathbb{N}
\ ,
\end{equation}
which, from Eq.~\eqref{R}, yields the mass function
\begin{equation}
\label{M}
m
=
M
-
\frac{Q^2}{2\,r}
+
\frac{1}{2}\,\sum_{n=0}^{\infty}\,\frac{C_n\,r^{n+1}}{(n+1)(n+2)}
\ ,
\end{equation}
for $0<r\leq\,h$, where $M$ and $Q$ are integration constants that can be identified with the mass of the
Schwarzschild solution and a charge for the Reissner-Nordstr\"{o}m (RN) geometry, respectively. 
However, since it is known that the RN geometry contains a Cauchy horizon, we impose $Q=0$.
This leaves us with the two charges $M$ and ${\cal M}$. 
\par
Let us notice that the series~\eqref{M} converges around $r=h$ as soon as we impose the
condition~\eqref{cond1}, but it remains to see if it can represent an analytic function in its
whole domain $0<r\leq\,h$.
Moreover, the Schwarzschild metric is simply given by the condition in Eq.~\eqref{mcond}, that is
$M={\cal M}\neq 0$ and $Q=C_n=0$ for all $n$ in Eq.~\eqref{M}.
In Ref.~\cite{Ovalle:2024wtv}, other interior solutions where found with
\begin{eqnarray}
\label{Schw-limit2}
M=Q=0
\ ,
\end{eqnarray}
which are determined by the total mass ${\cal M}$ and some of the $C_n\neq 0$, so that
the exterior is still given by the Schwarzschild solution in Eq.~\eqref{mtransform}.
From the mass function~\eqref{sources}, the energy density and pressures associated
with these interior geometries read 
\begin{eqnarray}
\label{energy}
\kappa\,\epsilon
&=&
\sum_{n=0}^{\infty}\,\frac{C_n\,r^{n-2}}{n+2}
=
-\kappa\,p_r
\\
\kappa\,p_t
&=&
-\frac{1}{2}\sum_{n=0}^{\infty}\,\frac{n}{n+2}\,C_n\,r^{n-2}
\ ,
\label{pt}
\end{eqnarray}
for $0<r\leq\,h$. 
\par 
The simplest of such solutions was found by imposing the continuity conditions~\eqref{cond2}
on the mass function~\eqref{M} [see Ref.~\cite{Ovalle:2024wtv} for all details], which yields
\begin{equation}
\label{m1}
m
=
r-\frac{r^2}{2\,h}
\ ,
\end{equation}
corresponding to the interior line element
\begin{equation}
\label{sol1}
ds^{2}
=
\left(1-\frac{r}{h}\right)\,dt^{2}
-
\frac{dr^2}{1-\frac{r}{h}}
+r^2\,d\Omega^2
\ ,
\quad
{\rm for}\ 
0< r\,{\color{blue}\leq}\,h.
\ 
\end{equation}
The source is given by
\begin{equation}
\label{sources1}
\kappa\,\epsilon
=
-\kappa\,p_r
=
\frac{2}{r^2}\left(1-\frac{r}{h}\right)
\ ,
\quad
\kappa\,p_\theta=\frac{1}{h\,r}
\ ,
\end{equation}
generating the curvature
\begin{equation}
\label{Rsin1}
R
=
\frac{4}{r^2}\left(1-\frac{3\,r}{2\,h}\right)
\ ,
\quad
{\rm for}\
0< r\,\,{\color{blue}\leq}\,\,h
\ .
\end{equation}
\par
A second solution can be found by imposing a smoother transition between the two regions
separated by the horizon, that is 
\begin{equation}
\label{cond3}
m''(h)
=
0
\ ,
\end{equation}
which yields
\begin{equation}
\label{M2}
m
=
r-\frac{r^3}{h^2}+\frac{r^4}{2\,h^3}
\ .
\end{equation}
The line element is
\begin{eqnarray}
\label{sol2}
ds^{2}
&=&
\left(1-\frac{2\,r^2}{h^2}+\frac{r^3}{h^3}\right)\,dt^{2}
-
\frac{dr^2}{1-\frac{2\,r^2}{h^2}+\frac{r^3}{h^3}}
\nonumber
\\
&&
+
r^2\,d\Omega^2
\ ,
\quad
{\rm for}\
0< r\,\,{\color{blue}\leq}\,\,h
\ ,
\end{eqnarray}
sourced by
\begin{eqnarray}
\label{sources2}
\kappa\,\epsilon
&=&
-\kappa\,p_r
=
\frac{2}{r^2\,h^3}\left(h-r\right)^2(h+2\,r)
\ ,
\nonumber
\\
\kappa\,p_\theta
&=&
\frac{6}{h^3}\left(h-r\right)
\ ,
\end{eqnarray} 
which produces a curvature 
\begin{equation}
\label{Rsin2}
R
=
\frac{4}{r^2}\left(1+\frac{5\,r^3}{h^3}-\frac{6\,r^2}{h^2}\right)
\ ,
\quad
{\rm for}\
0< r\,\,{\color{blue}\leq}\,\,h
\ .
\end{equation}
Finally, it can be proven~\cite{Ovalle:2024wtv} that the mass function~\eqref{M2}
is a particular case of
\begin{equation}
\label{Mfrac}
m
=
r
+
\frac{r}{n^2-2\,n-1}
\left(\frac{r}{h}\right)^n
\left[
1-
\frac{\left(n-1\right)^2}{2}
\left(\frac{r}{h}\right)^{\frac{2}{n-1}}\right]
\ ,
\end{equation}
where $n>1\in\mathbb{N}$ includes the polynomial case $n=2$ in Eq.~\eqref{M2}.
The mass function~\eqref{Mfrac} yields the metric function
\begin{equation}
\label{sol2n}
f
=
1
+
\left(\frac{r}{h}\right)^n
\frac{\left[2-\left(n-1\right)^2\left({r}/{h}\right)^{\frac{2}{n-1}}\right]}{n^2-2\,n-1}
\ .
\end{equation}
\par
It is easy to show that the BHs in Eqs.~\eqref{sol1}, \eqref{sol2} and~\eqref{sol2n}
have no inner horizon.
Indeed, as conjectured in Ref.~\cite{Ovalle:2024wtv}, apart from the Schwarzschild BH,
the simplest two single horizon BH solutions, with the total mass ${\cal M}$ as a unique
charge, are those displayed in Eqs.~\eqref{sol1} and~\eqref{sol2} for the region $0<r<h$,
which smoothly join the Schwarzschild exterior at the horizon $r=h=2\,{\cal M}$. %
\begin{figure}
\includegraphics[width=0.45\textwidth]{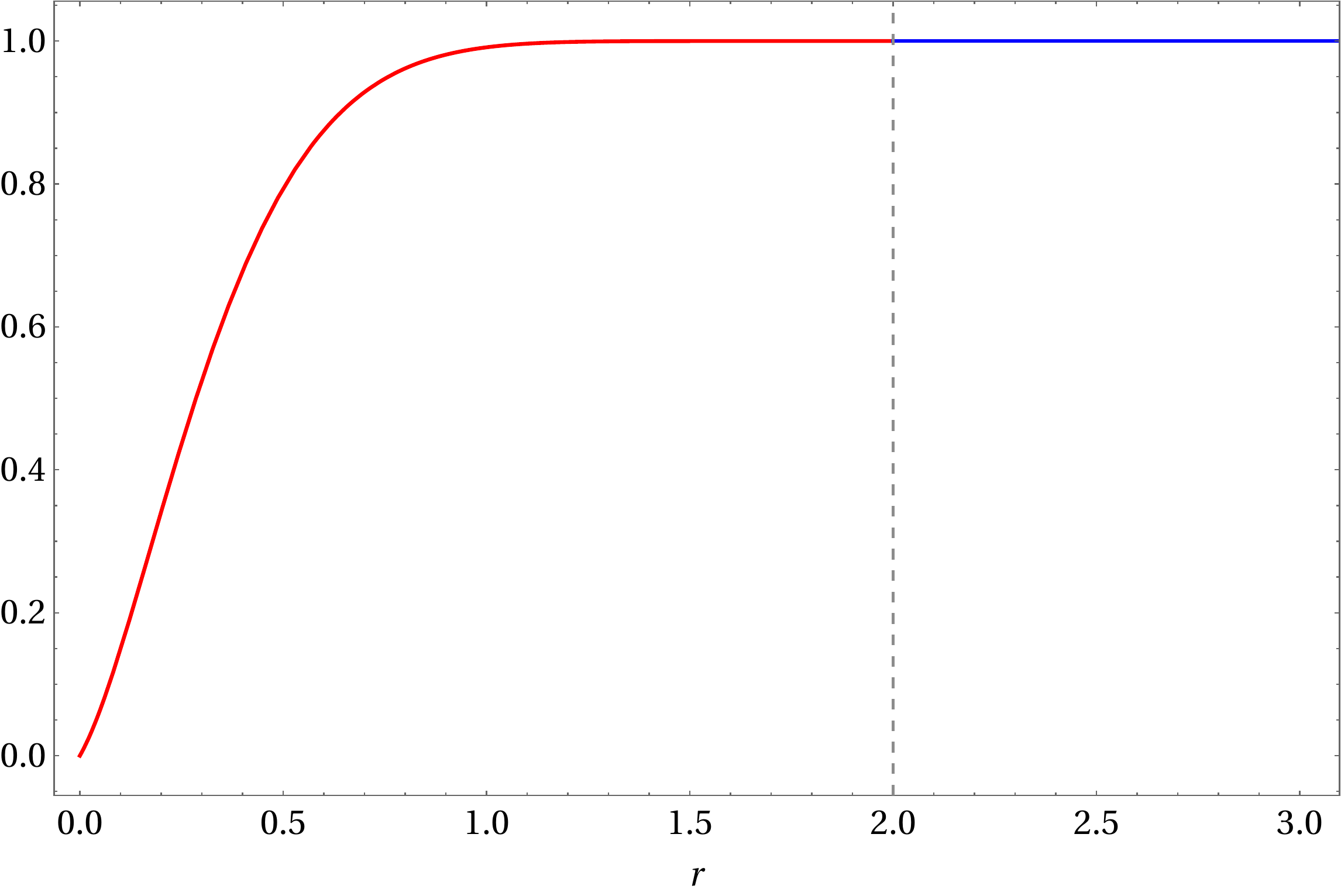}
\\
$\ $
\\
\includegraphics[width=0.45\textwidth]{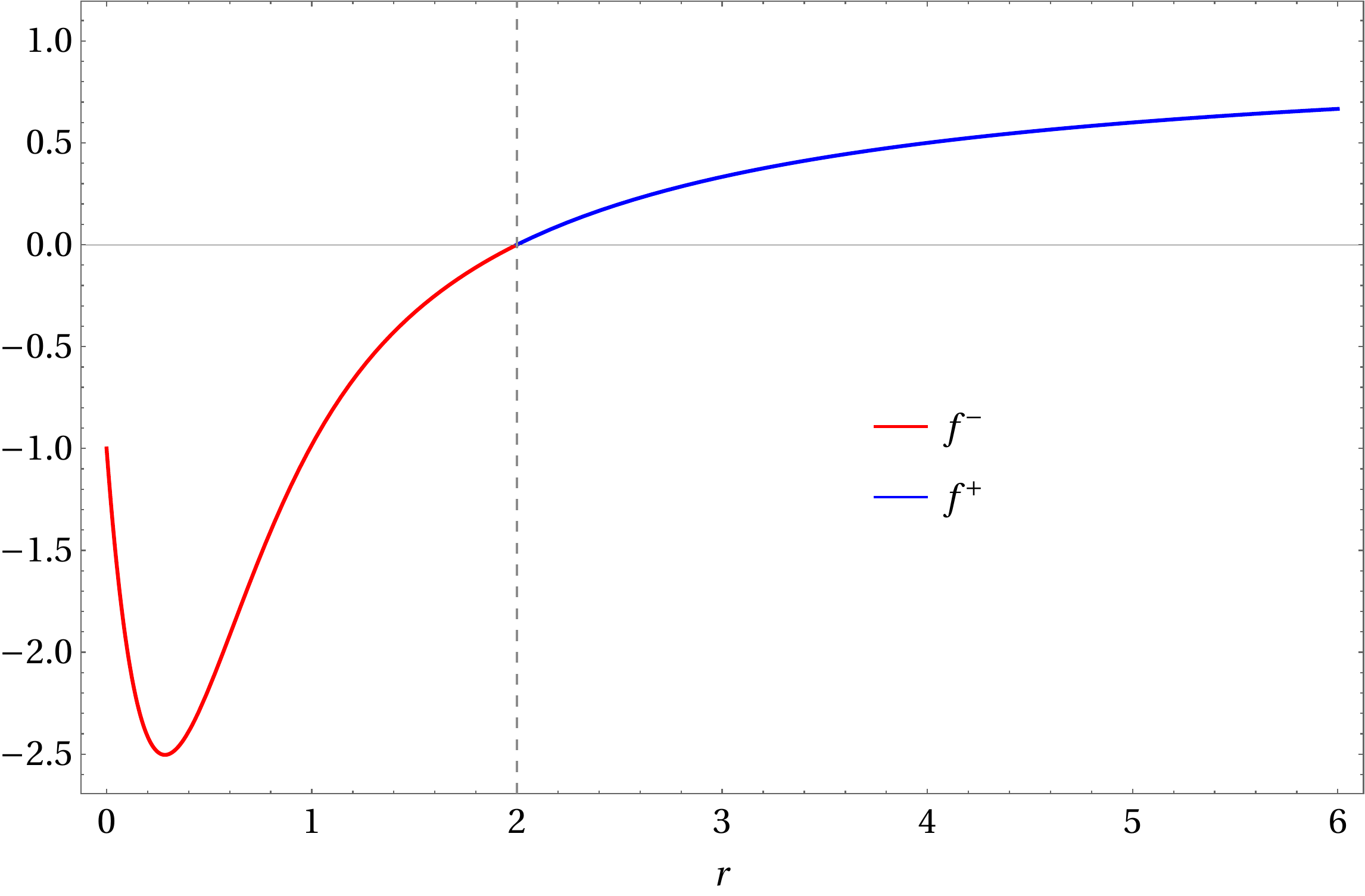}
\caption{Mass function in Eq.~\eqref{m10} (upper panel) and
corresponding metric function $f=1-2\,m/r$ (lower panel) for $N=10$.
The vertical line represents the horizon $h=2\,{\cal M}$ for ${\cal M}=1$
and red (blue) colour is for the interior (exterior).}
\label{fig1}
\end{figure}
\par
We can obtain more solutions by considering the metric function
in Eq.~\eqref{M} and truncating the series as~\footnote{Expressions in
Eqs.~\eqref{m1} and~\eqref{M2} correspond to~\eqref{mseries} for $N=3$ and $N=4$,
respectively.}
\begin{equation}
\label{mseries}
m
=
r
+
\sum_{i=2}^{N}\,C_i\,r^i
\ ,
\end{equation}
where $N>2$ and the $(N-1)$ unknown $C_i$ can be found by the condition~\eqref{cond1} and 
\begin{equation}
\label{cond-n}
\frac{d^n m}{dr^n}(h)
=
0
\ ,
\end{equation}
for all $n\leq N-2$. A straightforward consequence of the differential constraints~\eqref{cond-n}
is that the energy-momentum tensor is continuous at the horizon for $N>3$, that is,
\begin{equation}
T^{\mu}_{\ \nu}(h)
=0
\ .
\end{equation}
For example, the solution with $N=10$ is given by
\begin{eqnarray}
\label{m10}
m
&=&
r
+\frac{27\,r^2}{2\,h}-\frac{84\,r^3}{h^2}+\frac{231\,r^4}{h^3}-\frac{378\,r^5}{h^4}+\frac{399\,r^6}{h^5}
\nonumber
\\
&&
-\frac{276\,r^7}{h^6}+\frac{243\,r^8}{2\,h^7}-\frac{31\,r^9}{h^8}+\frac{7\,r^{10}}{2\,h^9}
\ ,
\end{eqnarray}
which is displayed in Fig.~\ref{fig1}.
A simple analysis of the expression for the mass function~\eqref{mseries} shows
that the strong energy condition is violated for $N>4$ ($p_\theta<0$ for $r\gtrsim 0$).
However, the weak energy condition still holds.
If we instead consider a polynomial solution of the form
\begin{equation}
\label{mpoly}
m
=
r
+
A\,r^l
+B\,r^n
+C\,r^p
\ ,
\quad
p\neq n\neq l>1
\ ,
\end{equation}
where $A$, $B$ and $C$ are constants to be determined from Eqs.~\eqref{cond2}
and~\eqref{cond3}, we generate the solutions displayed in Table~\ref{tab1}.
Indeed, we could go further by including additional terms in~\eqref{mpoly},
or just by relaxing the energy conditions, which are most likely violated at high
curvature.
Therefore, we can safely conclude that the inner region is much richer than
illustrated in Table~\ref{tab1}.
This will be particularly important for the cosmological models, as we will see next.
\begin{table*}
\caption{Interior geometries with mass function~\eqref{mpoly} satisfying $m'(h)=m''(h)=0$.
\label{tab1}}
\begin{ruledtabular}
\begin{tabular}{ c c c c }
$\{l,\,n,\,p\}$ & $m(r)=r+A\,r^l+B\,r^n+C\,r^p$ &$\epsilon>0$ & Energy condition
\\  \hline
$\{2,\,n,\,p\}$
&
$m(r)=r-\frac{[2-2p+n (p-2)]}{2 (n-2)(p-2)}\frac{r^2}{h}+\frac{h}{(n-2) (n-p)}\left(\frac{r}{h}\right)^n+\frac{h}{(p-2) (p-n)}
\left(\frac{r}{h}\right)^p
$ &Yes &Strong
\\  \hline
$\{3,\,4,\,p\}$ & $m(r)=r-\frac{r^3}{h^2}+\frac{r^4}{2h^3}$ &Yes &Strong 
\\  \hline
$\{3,\,5,\,p\}$&$m(r)=r-\frac{h}{4}\frac{(3 p-8)}{(p-3)}\left(\frac{r}{h}\right)^3+\frac{h}{4}\frac{(p-4)}
{(p-5)}\left(\frac{r}{h}\right)^5-\frac{h/2}{(p-3)(p-5)}\left(\frac{r}{h}\right)^p
$ &Yes ($6\leq p\leq 16$) &Strong
\\ \hline
$\{3,\,6,\,p\}$&$m(r)=r-\frac{h}{3}\frac{(2 p-5)}{(p-3)}\left(\frac{r}{h}\right)^3+\frac{h}{6}\frac{(p-4)}{(p-6)}\left(\frac{r}{h}\right)^6
-\frac{h}{(p-3)(p-6)}\left(\frac{r}{h}\right)^p$
&Yes ($7\leq p\leq 10$)
&Strong
\\ \hline
$\{3,\,7,\,8\}$&$m(r)=r-\frac{7 r^3}{10 h^2}+\frac{r^7}{2h^6}-\frac{3 r^8}{10h^7}$
&Yes
&Strong
\\ \hline
$\{4,\,5,\,6\}$&$m(r)=r-\frac{5 r^4}{2h^3}+\frac{3	r^5}{h^4}-\frac{r^6}{h^5}$
&Yes
&Strong
\end{tabular}
\end{ruledtabular}
\end{table*}
\begin{table*}
\caption{Cosmological form~\eqref{patch2} of the geometries in Table~\ref{tab1} with $t\leq t_0=h$ where
$\mu'(t_0)=1$ and $\mu''(t_0)=0$.
\label{tab2}}
\begin{ruledtabular}
\begin{tabular}{ c c c c }
$\{l,\,n,\,p\}$ & $F(t)=1-\frac{2\mu(t)}{t}\geq\,0\ .$ &$\epsilon>0$ &Energy condition
\\  \hline
$\{2,\,n,\,p\}$&$F(t)=1-\frac{[2-2p+n (p-2)]}{(n-2)(p-2)}\left(\frac{t}{t_0}\right)+\frac{2}{(n-2) (n-p)}\left(\frac{t}{t_0}\right)^{n-1}+\frac{2}{(p-2) (p-n)}\left(\frac{t}{t_0}\right)^{p-1}$
&Yes&Strong
\\  \hline
$\{3,\,4,\,p\}$ & $F(t)=1-2\left(\frac{t}{t_0}\right)^2+\left(\frac{t}{t_0}\right)^3$
&Yes&Strong 
\\  \hline
$\{3,\,5,\,p\}$&$F(t)=1-\frac{1}{2}\frac{(3 p-8)}{(p-3)}\left(\frac{t}{t_0}\right)^2+\frac{1}{2}\frac{(p-4)}{(p-5)}\left(\frac{t}{t_0}\right)^4-\frac{1}{(p-3)(p-5)}\left(\frac{t}{t_0}\right)^{p-1}
$&Yes ($6\leq p\leq 16$)&Strong
\\ \hline
$\{3,\,6,\,p\}$&$F(t)=1-\frac{2}{3}\frac{(2 p-5)}{(p-3)}\left(\frac{t}{t_0}\right)^2
+\frac{1}{3}\frac{(p-4)}{(p-6)}\left(\frac{t}{t_0}\right)^5-\frac{2}{(p-3)(p-6)}\left(\frac{t}{t_0}\right)^{p-1}$
&Yes ($7\leq p\leq 10$)
&Strong
\\ \hline
$\{3,\,7,\,8\}$&$F(t)=1-\frac{7}{5}\left(\frac{t}{t_0}\right)^2+\left(\frac{t}{t_0}\right)^6-\frac{3}{5}\left(\frac{t}{t_0}\right)^7$
&Yes
&Strong
\\ \hline
$\{4,\,5,\,6\}$&$F(t)=1-5\,\left(\frac{t}{t_0}\right)^3+6\,\left(\frac{t}{t_0}\right)^4-2\,\left(\frac{t}{t_0}\right)^5$
&Yes
&Strong
\end{tabular}
\end{ruledtabular}
\end{table*}
\section{Cosmology}
\label{sec4}
All of the interior BH solutions in Table~\ref{tab1} can be considered as a whole
universe~\cite{Doran:2006dq}, which is precisely what we will explore next.
Let us start by noticing that for $0<r\le h$ the line element has the form
\begin{eqnarray}
\label{patch1}
ds^{2}
=
F(r)\,dt^{2}
-\frac{dr^2}{F(r)}
+r^2\,d\Omega^2
\ ,
\end{eqnarray}
where
\begin{equation}
F
=
1-\frac{2\,\mu(r)}{r}
\geq
0
\ ,
\end{equation}
and
\begin{equation}
\mu
=
-\left(A\,r^l+B\,r^n+C\,r^p\right)
\end{equation}
can be read directly from Table~\ref{tab1}.
We can rewrite the metric~\eqref{patch1} by making explicit the role of time
and radial coordinates as
\begin{eqnarray}
\label{patch2}
ds^{2}
=
-\frac{dt^2}{F(t)}
+
F(t)\,dr^{2}
+t^2\,d\Omega^2
\ ,
\end{eqnarray}
where
\begin{equation}
\label{F}
F
=
1-\frac{2\mu(t)}{t}\geq\,0
\end{equation}
is displayed in Table~\ref{tab2} for each case of Table~\ref{tab1}, respectively,
with $0<t<t_0=h$.
\par
We can further write the metric~\eqref{patch2} in terms of the cosmic 
(or synchronous) time defined by
\begin{equation}
\label{tau}
d\tau=\pm\frac{dt}{\sqrt{F(t)}}
\ ,
\end{equation}
which leads to the generic cosmological solution
\begin{equation}
\label{cosmo}
ds^2
=
-d\tau^2
+a^2(\tau)\,dr^{2}
+b^2(\tau)\,d\Omega^2
\ .
\end{equation}
The metric~\eqref{cosmo} represents a Kantowski-Sachs
universe~\cite{Kantowski:1966te,Brehme:1977fi} with the two
scale factors
 \begin{eqnarray}
\label{scalefactors}
a^2(\tau)
&\equiv&
F(\tau)
\nonumber
\\ 
\\
b^2(\tau)
&\equiv&
t^2(\tau)
\ .
\nonumber
\end{eqnarray}
This solution in general describes an homogeneous but anisotropic universe,
with Einstein tensor 
\begin{eqnarray}
\label{einsteintensor}
G^0_{\ 0}
&=&
-\left(\frac{1}{b^2}
+\frac{2\,\dot{a}\,\dot{b}}{a\,b}
+\frac{\dot{b}^2}{b^2}\right)
\\ 
G^1_{\ 1}
&=&
-\left(\frac{1}{b^2}+\frac{2\,\ddot{b}}{b}+\frac{\dot{b}^2}{b^2}\right)
\\ 
G^2_{\ 2}
&=&
-\left(\frac{\dot{a}\,\dot{b}}{a\,b}+\frac{\ddot{a}}{a}+\frac{\ddot{b}}{b}\right)
\ .
\end{eqnarray}
\par
Let us consider, for instance, the simplest inner BH given by the
metric~\eqref{sol1}, which yields
\begin{equation}
\label{cosmo1}
ds^2
=
-\frac{dt^2}{1-{t}/{t_0}}
+
\left(1-\frac{t}{t_0}\right)dr^2
+
t^2\,d\Omega^2
\ ,
\end{equation}
for $0<t<t_0$.
We have
\begin{equation}
F
=1-\frac{t}{t_0}
\ ,
\end{equation}
which leads to the cosmic time
\begin{equation}
\label{tau2}
\tau(t)
=
-2\,t_0\,\sqrt{1-\frac{t}{t_0}}
\ ,
\quad
{\rm for}\
t<t_0
\ ,
\end{equation}
and
\begin{eqnarray}
\label{cosmo1f}
ds^{2}
=
-d\tau^2
+
\frac{\tau^2}{\tau_0^2}\,dr^{2}
+
\frac{\tau_0^2}{4}
\left(\frac{\tau^2}{\tau_0^2}-1\right)^2d\Omega^2
\ ,
\end{eqnarray}
where $\tau_0\equiv 2\,t_0$.
Notice that in this case the scale factors satisfy
\begin{equation}
b^2
=
\frac{\tau_0^2}{4}\left(a^2-1\right)^2
\ .
\end{equation}
\par 
The source of the metric~\eqref{cosmo1f} is given by
 \begin{eqnarray}
\label{sources1Cosmo}
\kappa\,\epsilon
&=&
-\kappa\,p_r
=
\frac{8\,\tau^2}{\left(\tau^2-\tau_0^2\right)^2}
\\
\kappa\,p_\theta
&=&
\frac{4}{\tau_0^2-\tau^2}
\ ,
\end{eqnarray}
with curvature
\begin{equation}
\label{Rsin1Cosmo}
R
=
\frac{8\,\left(3\,\tau^2-\tau_0^2\right)}{\left(\tau^2-\tau_0^2\right)^2}
\ .
\end{equation}
The anisotropy for this example is therefore given by
\begin{equation}
\Delta
\equiv
p_\theta-p_r
=
\frac{4}{\kappa}\frac{\tau^2+\tau_0^2}{\left(\tau^2-\tau_0^2\right)^2}
\ .
\end{equation}
\par
The above example contains a curvature singularity in Eq.~\eqref{Rsin1Cosmo}
for $\tau\to\tau_0$.
In fact, we can study the behaviour of these universes in the vicinity of the
cosmological singularity in general. 
The curvature scalar of the metric~\eqref{patch2} is given by
\begin{equation}
R
=
{F''}+\frac{4\,{F'}}{t} + \frac{2\,F}{t^2}+\frac{2}{t^2}
\ ,
\label{curvature}
\end{equation}
where primes stand for derivatives with respect to $t$.
Since all of the functions $F$ in Table~\ref{tab2} are polynomials in $t$
with the constant term equal to $1$, Eq.~\eqref{curvature} is singular at $t=0$.
In fact, it is easy to see that the curvature behaves as 
\begin{equation}
R
\approx
\frac{4}{t^2}
\ ,
\quad
{\rm for}\ t\to 0
\ ,
\label{singt}
\end{equation}
for all the functions $F$.
\par
We can also consider the cosmic time 
\begin{equation}
d\tau
=
\frac{dt}{\sqrt{F(t)}}
\ .
\label{time}
\end{equation}
The function $F(t)\approx 1$ in the vicinity of the singularity $t=0$ and,
with a convenient choice of the integration constant, we have 
\begin{equation}
\tau
\approx
t
\ .
\label{time1}
\end{equation}
Thus, the curvature singularity in Eq.~\eqref{singt} can also be written as 
\begin{equation}
R
\approx
\frac{4}{\tau^2}
\ ,
\quad
{\rm for}\ \tau\to 0
\ .
\label{sing}
\end{equation}
\par
It is interesting to compare this result with the singularities arising in
isotropic Friedmann cosmologies.
Let us consider a flat Friedmann universe with the metric
\begin{equation}
ds^2
=
-d\tau^2
+a^2(\tau)
\left(dr^2+r^2\,d\Omega^2\right)
\ ,
\label{Fried}
\end{equation}
whose scalar curvature is given by
\begin{equation}
R
=
6\left(\frac{\ddot{a}}{a}
+
\frac{\dot{a}^2}{a^2}
\right)
\ ,
\label{Fried1}
\end{equation}
where dots stand for derivatives with respect to $\tau$.
For a power law expansion,
\begin{equation}
a
=
a_0\,\tau^{k}
\ ,
\label{Fried2}
\end{equation}
we have   
\begin{equation}
R
=
\frac{6\,k\,(2\,k-1)}{\tau^2}
\ .
\label{Fried3}
\end{equation}
This leads to the same singularity as the one in Eq.~\eqref{sing} if
$6\,k\,(2\,k-1)=4$, or
\begin{equation}
k
=
\frac14\left(1+\sqrt{\frac{19}{3}}\right)
\ .
\label{k}
\end{equation}
It is known that a {homogeneous and isotropic}
universe which expands according to the power law~\eqref{Fried2}
is filled with a barotropic fluid with equation of state 
\begin{equation}
p
=
w\,\varepsilon
\ ,
\label{w}
\end{equation}
where the parameter $w$ is constant and 
\begin{equation}
k
=
\frac{2}{3\,(1+w)}
\ ,
\label{k1}
\end{equation}
or
\begin{equation}
w
=
\frac{2}{3\,k}-1
\ .
\label{w1}
\end{equation}
From Eq.~\eqref{k}, we then find
\begin{equation}
w
=
\frac{8}{3\left(1+\sqrt{{19}/{3}}\right)} -1
\simeq
-0.24
\ .
\label{w2}
\end{equation}
This means that {our homogeneous but anisotropic universe behaves
near the singularity like a homogeneous and isotropic universe driven by a isotropic}
fluid with negative pressure but equation of state parameter
$w>-1/3$, which is the critical value below which the deceleration would turn into
acceleration. 
\section{Conclusion}
\label{sec5}
Generating cosmological models from BH geometries is a well-known
procedure which, in general, allows us to develop solutions beyond the
standard (isotropic and homogeneous) cosmological models.
On the other hand, a plethora of new BH solutions have been reported
in recent years, whose interest is especially due to their interpretation
in terms of nonlinear electrodynamics~\cite{Ayon-Beato:1998hmi}.
This leads to the possibility of constructing a plethora of new cosmological
models as well.
Obviously, this could become counterproductive if what we seek are
cosmological alternatives, beyond the standard model, whose origin is fully
justified by first principles.
\par
In this sense, the advantage of our cosmological solutions in
Table~\ref{tab2} is that they are derived from a family of very non-trivial
BH geometries.
They are solutions that in fact tell us a lot about how complex the interior
of the simplest spherically symmetric BHs could be.
Therefore, their cosmological versions, as well as possible extensions,
are quite attractive, especially if we want to justify processes that are
not yet well understood, such as the phenomenology of dark matter
and dark energy, and possible explanations based on cosmological
models other than the presently dominant one.
\subsection*{Acknowledgments}
R.C.~and A.K.~are partially supported by the INFN grant FLAG.
The work of R.C.~has also been carried out in the framework of activities of the
National Group of Mathematical Physics (GNFM, INdAM).
J.O.~is partially supported by ANID FONDECYT~Grant No.~1210041.
%
%
%
%
\bibliography{references.bib}
\bibliographystyle{apsrev4-1.bst}
%
%
\end{document}